\documentclass[a4paper, oneside, twocolumn, notitlepage, 10pt]{extarticle_ecoc}
\usepackage{ecoc}
\usepackage{subfigure}
\begin{document}
\selectlanguage{english}    


\title{End-to-end Learning of a Constellation Shape Robust to Variations in SNR and Laser Linewidth}%


\author{
    Ognjen Jovanovic, 
    Metodi P. Yankov, 
    Francesco Da Ros, 
    Darko Zibar 
}

\maketitle                  


\begin{strip}
 \begin{author_descr}

   Department of Photonics Engineering, Technical University of Denmark,
   \textcolor{blue}{\uline{ognjo@fotonik.dtu.dk}} 



 \end{author_descr}
\end{strip}

\setstretch{1.1}
\renewcommand\footnotemark{}
\renewcommand\footnoterule{}
\let\thefootnote\relax\footnotetext{978-1-6654-3868-1/21/\$31.00 \textcopyright 2021 IEEE}

\begin{strip}
  \begin{ecoc_abstract}
    We propose an autoencoder-based geometric shaping that learns a constellation robust to SNR and laser linewidth estimation errors. This constellation maintains shaping gain in mutual information (up to $0.3$~bits/symbol) with respect to QAM over various SNR and laser linewidth values.
  \end{ecoc_abstract}
\end{strip}


\section{Introduction}
Geometric constellation shaping (GCS) is used to optimize high-order modulation formats to improve the spectral efficiency and maximize mutual information (MI). For coherent optical communication systems, such optimization should include residual phase noise (RPN) which results from an imperfect carrier phase estimation (CPE) and compensation. The parametrization of CPE algorithms, such as the ubiquitous blind phase search (BPS)\supercite{Pfau2009a}, is sensitive to the channel conditions, such as signal-to-noise ratio (SNR) and laser linewidth (LW). In practical scenarios, measuring the laser LW is challenging and LW may drift over time, e.g. due to aging. Interoperatibility between vendors is becoming an increasingly important characteristic of optical networks\supercite{Filer:18}, which means that the transmission needs to support a variety of hardware with different components, resulting in varying SNR and laser LW. It is therefore of the utmost importance to find a constellation that maintains good performance under imperfect knowledge of the channel conditions.

Performing GCS, which usually relies on gradient-based optimization, on a channel model that includes the PN and the CPE could be challenging because the CPE is usually complex and non-differentiable, e.g. the BPS algorithm. Therefore, previous works on GCS assume ideal knowledge of channel conditions and artificially modeled RPN \supercite{Li,Pfau,Krishnan,Kayhan,Dzieciol}. However, this assumption does not reflect the true RPN after the CPE which is often mis-parametrized due to imperfect knowledge of the channel conditions.



In this paper, an autoencoder (AE) is used to geometrically optimize a constellation that is robust to variations in SNR and LW. The robust constellation was learned by varying the RPN severity and SNR in a simple differentiable RPN channel in the training stage. The constellation is then tested on a realistic channel with BPS, where the RPN is due to the mismatch between the channel conditions and the BPS parameters. Up to $0.3$~bits/symbol of shaping gain in MI with respect to QAM are achieved for a various degree of channel conditions mismatch in terms of LW and SNR.

\section{Autoencoder and channel models}

\begin{figure*}[t]
   \centering
    \includegraphics[width=0.85\linewidth]{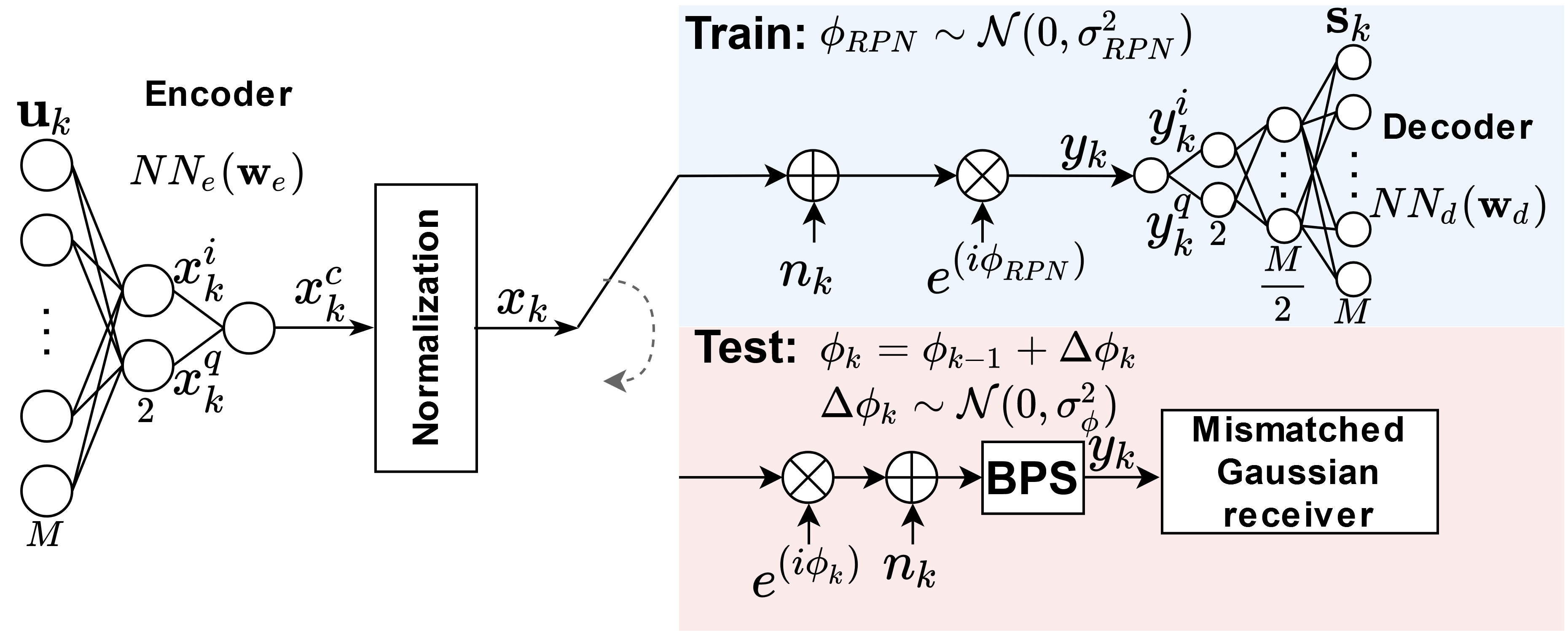}
    \caption{The training and testing setup of the autoencoder model used for geometrical constellation shaping.}
    \label{fig:Setup}
\end{figure*}

An AE, which consist of an encoder, a decoder, and an embedded differentiable channel model in between, is utilized to geometrically optimize a constellation \supercite{Jones2018a}, as shown on Fig. \ref{fig:Setup}. Two different setups can be distinguished: the training and the testing setup. Both of the setups share the same encoder. In the training setup the channel model is a simple differentiable approximation of the test channel which is more suitable for training.

The encoder, which learns the geometrically optimized constellation, is represented by a linear feed-forward neural network (NN) $NN_e(\mathbf{w_e})$ parameterized with trainable weights $\mathbf{w_e}$. It performs a mapping of the one-hot encoded vector $\mathbf{u}_k \in \mathbb{U} = \{ \mathbf{e}_i | i=1,\dots, M\}$ to a normalized complex constellation point $x_k$, where $k$ represents the $k$-th sample, $\mathbf{e}_i$ is an all zero vector with a one at position $i$ and $M$ is the constellation size. 

In the training setup, the channel consists of complex AWGN and RPN, both modelled as zero-mean Gaussian distributions with variance $\sigma^2_{n}$ and $\sigma^2_{RPN}$, respectively. A decoder NN $NN_d(\mathbf{w_d})$ with trainable weights $\mathbf{w_d}$ and using a softmax output layer is used as a receiver during training. The decoder's goal is to reproduce the input sequence $\mathbf{u}_k$ at the output $\mathbf{s}_k$ with the highest fidelity. This is achieved by jointly optimizing the encoder and the decoder trainable weights. The optimization of these weights is performed by minimizing the cross-entropy cost function such that $\mathbf{s}_k \approx \mathbf{u}_k$. Once the training has converged, the encoder weights are fixed and the testing is performed. 

In the testing setup, the channel consists of phase noise modelled as a Weiner process with variance $\sigma^2_\phi$, complex AWGN modelled as zero-mean Gaussian distribution with variance $\sigma^2_{n}$ and BPS as the blind CPE algorithm. The BPS algorithm is parametrized by the number of test phases $N_s=60$ and window size $W=128$ which were chosen so that the non-shaped QAM constellation performs well on average across the studied SNR and LW conditions. During testing, the decoder is replaced by the common mismatched Gaussian receiver\supercite{Dzieciol} to estimate the MI between the channel input and output in order to study the performance of the constellation.

In both setups, the channels operate at one sample per symbol with a symbol rate $R_s=32$~GBd. The AWGN variance is determined by the SNR, $\sigma^2_{n}=\frac{1}{SNR}$. The PN process variance $\sigma^2_\phi$ is determined by the laser LW $\Delta\nu$ and symbol period $T_s = 1/R_s$,  $\sigma^2_\phi = 2\pi\Delta\nu T_s$. The constellation size is $M=64$. The AE hyperparameters are shown in Table \ref{tab:Setup}.

\begin{table}[t]
   \centering
\caption{Autoencoder hyperparameters}
\label{tab:Setup}%
      \begin{tabular}{|c|c|c|}
        \hline  & Encoder NN & Decoder NN \\
        \hline \# of input nodes & M & 2\\
        \hline \# of hidden layers & 0 & 1\\
        \hline \# of hidden nodes & 0 & $M/2$\\ 
        \hline \# of output nodes & 2 & M\\
        \hline Bias & No & Yes\\
        \hline Hidden layer & None & Leaky Relu\\
        activation function & & \\
        \hline Output layer & Linear & Softmax\\
        activation function & & \\
        \hline
      \end{tabular}
\end{table}

\section{Results and Discussion}

\begin{figure}[t]
   \centering
    \includegraphics[width=0.9\linewidth]{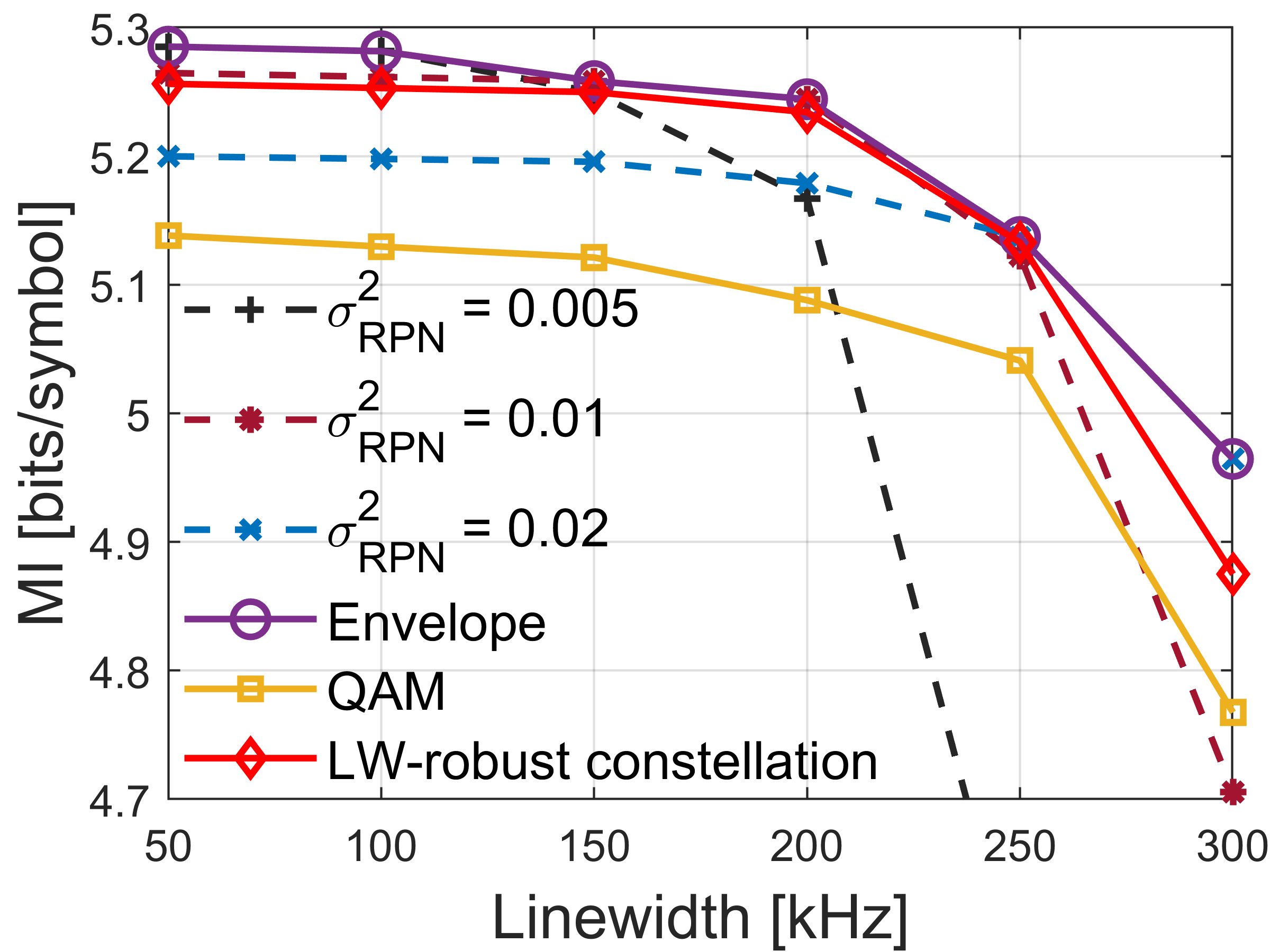}
    \caption{Performance in MI with respect to LW for the constellation trained by varying $\sigma_{RPN}^2$ at SNR~$=17$~dB.}
    \label{fig:LW_robust_MI}
\end{figure}

\begin{figure}[t]
   \centering
        \includegraphics[width=0.9\linewidth]{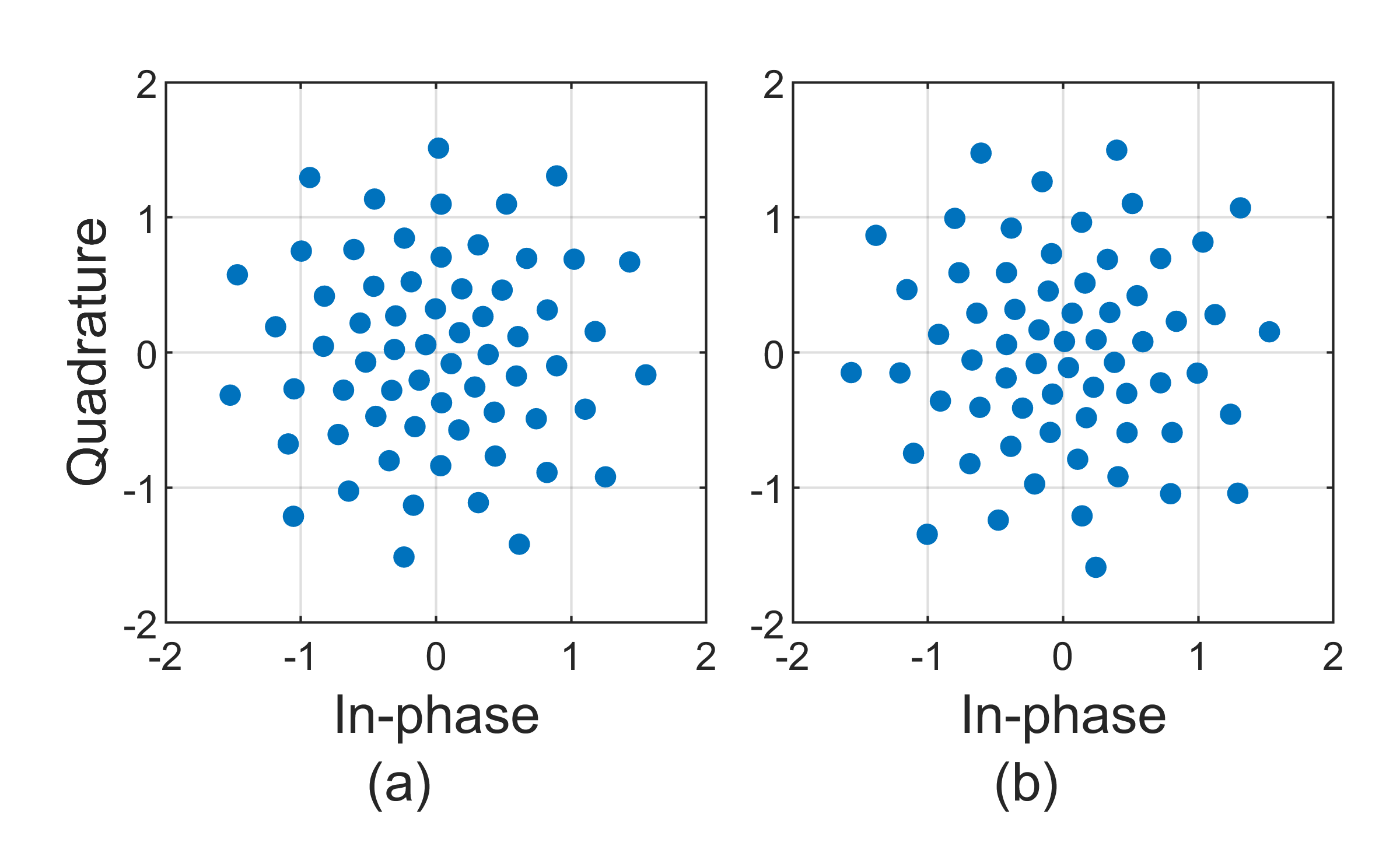}
    \caption{Constellation robust to (a) varying LW for a fixed SNR$=17$~dB; (b) varying SNR and LW.}
    \label{fig:robust_constellations}
\end{figure}

\begin{figure*}[t]
   \centering
    \includegraphics[width=0.9\linewidth]{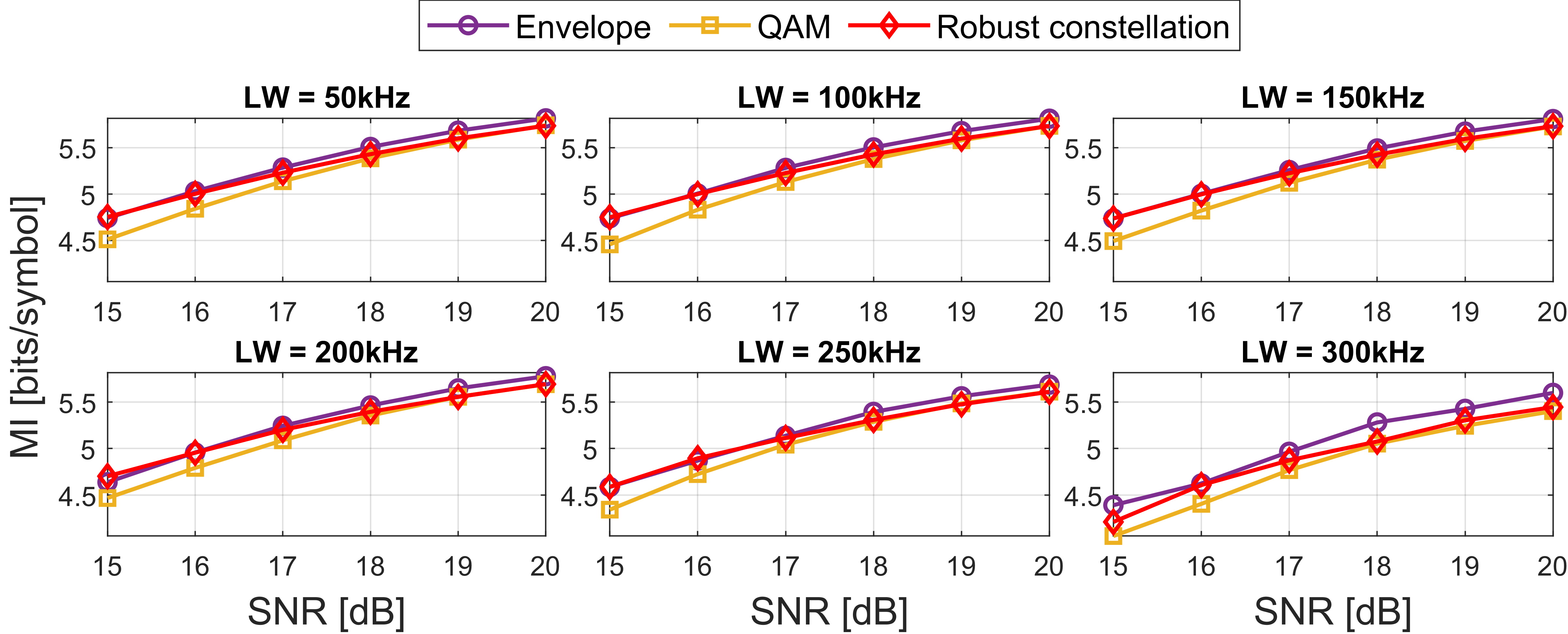}
    \caption{Performance in MI with respect to SNR for LW values for the constellation trained by varying both $\sigma_{RPN}^2$ and SNR.}
    \label{fig:LW_SNR_robust_MI}
\end{figure*}



We compare our learned robust constellation to two cases: 1) a conventional square quadrature amplitude modulation (QAM); 2) constellations trained on fixed SNR and RPN variance $\sigma^2_{RPN}$, similar to what was done in \supercite{Dzieciol}. The RPN variance is taken from a coarsely chosen set $\sigma^2_{RPN} \in \{10^{-4},5 \cdot 10^{-4},10^{-3},5 \cdot 10^{-3},10^{-2},2 \cdot 10^{-2},5 \cdot 10^{-2}\}$. Case 2) approaches the best performing constellation with regards to MI for a given SNR and LW pair and assumes they are known perfectly at both the transmitter and receiver. In both the training and the testing stage the studied SNR region includes values from the interval SNR~$\in \{15,16,\dots,20\}$~dB. In the testing stage, the studied laser LWs are $\Delta \nu \in \{50,100,\dots, 300\}$~kHz.

The first goal was to achieve a constellation robust over various LWs for a target SNR (referred to as LW-robust constellation in the following). The training to learn such a constellation is exemplified by fixing the SNR to~$17$~dB and sampling the RPN variance from a log-uniform distribution in the range of $\sigma^2_{RPN} \in [0.005, 0.02]$. Then, in order to achieve a single constellation that is robust over all target SNR and LW pairs (referred to as SNR\&LW-robust constellation in the following.), the training was performed on uniformly distributed SNR $\in [15, 20]$~dB and log-uniformly distributed RPN variance $\sigma^2_{RPN} \in [0.005, 0.05]$. \emph{The SNR and RPN variance were drawn from these distributions for each training batch.} In both scenarios, the RPN variance range was chosen based on the minimum and maximum fixed RPN variance values that contribute to the envelope. The envelope represents the MI of the constellation at each SNR and LW pair obtained with the corresponding optimal $\sigma^2_{RPN}$. The SNR range in the second scenario was chosen to cover the whole target SNR region.

The training was done by applying the Adam optimizer\supercite{kingma2014adam} on a sample set of size $N=256 \cdot M$. In each epoch, a new sample set is generated with uniformly distributed one-hot encoded vectors and divided into 8 batches of size $B=32 \cdot M$. The testing was done by running $100$ simulations with $10^5$ symbols per simulation in each case.

The simulation results obtained from testing the LW-robust constellation for a fixed SNR~$=17$~dB and varying LW are shown in Fig. \ref{fig:LW_robust_MI}. Only the constellations trained with a fixed RPN variance that contribute to the envelope are shown. The AE trained with a fixed RPN is only beneficial in a limited range of LW. For example, when the LW is fixed at $\Delta \nu=100$~kHz, the RPN method from \cite{Dzieciol} can be used with $\sigma^2_{RPN}=0.005$ to achieve a potentially optimal constellation for this LW. However, for larger LWs this constellation becomes highly sub-optimal. Although the LW-robust constellation has a slight penalty compared to the envelope, it maintains MI gain compared to QAM, up to $0.15$~bits/symbol over the whole observed LW interval. The constellation is shown on Fig. \ref{fig:robust_constellations}(a).

Fig. \ref{fig:LW_SNR_robust_MI} shows MI performance of the envelope, QAM, and SNR\&LW-robust constellation as a function of SNR for different LW values. In this case, the SNR\&LW-robust constellation was trained by varying both SNR and RPN variance during training. The constellation is given in Fig. \ref{fig:robust_constellations}(b) and is used for all tests shown in Fig. \ref{fig:LW_SNR_robust_MI}. The SNR\&LW-robust constellation has a similar trend for all LW values. It achieves substantial gain at a SNR region from $15$ to $18$~dB, which is comparable to the constellation obtained with perfect knowledge of the channel conditions. The gain is then reduced at higher SNR, but the performance is still superior than regular QAM. The SNR\&LW-robust constellation achieves up to $0.3$~bits/symbols gain with respect to QAM for $\Delta \nu=100$~kHz, whereas the highest gain for the envelope is $0.33$~bits/symbol for $\Delta \nu=300$~kHz.


\section{Conclusions}
Autoencoder-based optimization of geometric shapes robust to variations in SNR and laser linewidth was proposed. The robustness of the constellation was achieved by utilizing a simpler channel model that includes additive white Gaussian noise and residual phase noise, and varying their severity for each batch of the training stage. This constellation maintains the shaping gain in mutual information with respect to QAM over the studied SNR and laser linewidth intervals in the testing phase which includes a realistic model of residual phase noise due to the BPS algorithm.



\section{Acknowledgements}
\small
This work was financially supported by the European Research Council through the ERC-CoG FRECOM project (grant agreement no. 771878), the Villum Young Investigator OPTIC-AI project (grant no. 29334), and DNRF SPOC, DNRF123.


\printbibliography

\end{document}